\documentclass[12pt]{article}
\usepackage{fullpage}

\usepackage{amsthm}
\usepackage{amsmath}
\usepackage{amsfonts}
\usepackage{graphicx}
\usepackage{url}

\newtheorem{thm}{Theorem}

\newtheorem{alg}[thm]{Algorithm}

\newtheorem{example}[thm]{Example}
\newtheorem{definition}[thm]{Definition}

\newcommand{\rr}{\mathbb{R}}
\newcommand{\e}{\mathrm{e}}
\newcommand{\tr}{\mathrm{tr}}
\newcommand{\bof}{\mathbf{f}}
\newcommand{\Ta}{T_1}
\newcommand{\Tb}{T_2}
\newcommand{\Tc}{T_3}
\newcommand{\hpt}{\hat{p}(\theta)}

\begin{document}
\title{Metric learning for phylogenetic invariants}
\author{Nicholas Eriksson\\
\url{nke@stanford.edu}\\
Department of Statistics,\\ 
Stanford University, \\
Stanford, CA 94305-4065
\and
Yuan Yao\\
\url{yuany@math.stanford.edu}\\
Department of Mathematics,\\ 
Stanford University, \\
Stanford, CA 94305-4065
}
\date{\today}

\maketitle

\begin{abstract}
We introduce new methods for phylogenetic tree quartet construction by using
machine learning to optimize the power of phylogenetic invariants.
Phylogenetic
invariants are polynomials in the joint probabilities which vanish under a
model of evolution on a phylogenetic tree.  We give algorithms for selecting a
good set of invariants and for learning a metric on this set of invariants
which optimally distinguishes the different models.  Our learning algorithms
involve linear and semidefinite programming on data simulated over a wide range
of parameters.  We provide extensive tests of the learned metrics on simulated data
from phylogenetic trees with four leaves under the Jukes-Cantor and Kimura
3-parameter models of DNA evolution.  Our method greatly improves on other uses
of invariants and is competitive with or better than neighbor-joining.  In
particular, we obtain metrics trained on trees with short internal branches
which perform much better than neighbor joining on this region of parameter
space.
\end{abstract}
\begin{quotation}
\noindent \small {\bf Keywords:} 
Phylogenetic invariants, algebraic statistics, semidefinite programming,
Felsenstein zone.  
\end{quotation}

\section{Introduction}

Phylogenetic invariants have been used for tree construction since the linear
invariants of Lake and Cavender-Felsenstein \cite{Lake1987,Cavender1987} were
found.  Although the linear invariants of the Jukes-Cantor model are powerful
enough to asymptotically distinguish between trees on 4 taxa \cite{Lake1987},
these linear invariants do not perform well in simulations
\cite{Huelsenbeck1995}.

In the last 20 years, the entire set of phylogenetic invariants has been found
for many models of evolution (see \cite{Allman2007} and the references
therein).  Since the invariants provide an essentially complete description of
the model, using more invariants should give more power to distinguish between
different models.  However, different invariants give vastly different power at
distinguishing models and it is not known how to find the most powerful
invariants.  

In this paper, we use techniques from machine learning to find metrics on the
space of invariants which optimize their tree reconstruction power for the
Jukes-Cantor and Kimura 3-parameter phylogenetic models on trees with four
leaves.  Specifically, we apply \emph{metric learning} algorithms inspired by
\cite{XingNg03} to find the metric which best distinguishes the models.  For
training data we use simulations over a wide range of the parameter space.  
Our main biological result is the construction of a metric which outperforms
neighbor-joining on trees simulated from the Felsenstein zone (i.e., trees with
a short interior edge). More generally, we find that metric learning
significantly improves upon other uses of invariants and is competitive with
neighbor joining even for short sequences and homogeneous rates.  

Casanellas and Fern\'{a}ndez-S\'{a}nchez \cite{Casanellas2006} also used the Kimura
3-parameter invariants to construct trees with four taxa.  Their results
indicated that invariants can sometimes perform better than commonly used methods (e.g.,
neighbor joining and maximum likelihood) for data that evolved with
non-homogeneous rates and for extremely long sequences.   They used the $l_1$
norm on the space of invariants, weighing each polynomial equally. 

This paper improves upon  \cite{Casanellas2006}
by showing how to improve upon the $l_1$ norm on the space of invariants.
The $l_1$ norm behaves poorly
since it weighs equally
informative and non-informative invariants.
Simulating data and using metric learning
improves the performance of invariants by putting much more weight on the
powerful invariants.  
This allows us to build an algorithm which is very accurate for trees with
short internal edges.



We begin by briefly introducing the models and phylogenetic invariants we will use.
Section~\ref{sec:methods} describes the metric learning algorithms; section~\ref{sec:results} gives
the results of our simulation studies; and we conclude with a short discussion.

By \emph{phylogenetic tree}, we mean a binary, unrooted tree with labelled
leaves.  There are three such trees with four leaves labelled $0, 1, 2, 3$, we
call these trees $\Ta$, $\Tb$, and $\Tc$ according to which leaf is on the same
``cherry'' as leaf 0.
We consider two phylogenetic models on these trees:  the Jukes-Cantor (JC69)
model of evolution \cite{Jukes1969} and the Kimura 3-parameter (K81) model
\cite{Kimura1981}, both with uniform root distribution.

These models associate to each edge $e$
of the tree a transition matrix 
$M_e = \e^{Qt_e}$ where 
where $t_e$ is the length of the edge $e$
and $Q$ is a rate matrix:
\[
Q = 
\begin{pmatrix}
	-3 \alpha & \alpha & \alpha & \alpha \\
	\alpha & -3 \alpha & \alpha & \alpha \\
	\alpha & \alpha & -3 \alpha & \alpha \\
	\alpha & \alpha & \alpha & -3 \alpha\\
\end{pmatrix}
\quad \text{ or } \quad 
\begin{pmatrix}
	\cdot & \gamma & \alpha & \beta\\
	\gamma & \cdot & \beta & \alpha\\
	\alpha & \beta & \cdot & \gamma\\
	\beta & \alpha & \gamma & \cdot
\end{pmatrix}
\]
for JC69 and K81 respectively, where $\cdot = -\gamma - \alpha - \beta$.
For a given tree, we write 
$p_{ijkl} = \Pr(\text{leaf } 0 = i, \text{leaf } 1 = j, \text{leaf }  2 = k, 
\text{leaf } 3 = l)$ for
$i,j,k,l \in \{\texttt{A,C,G,T}\}$ and write $p = (p_{\texttt{AAAA}}, \dots,
p_{\texttt{TTTT}})$ for the joint probability distribution. 
\emph{Phylogenetic invariants} are polynomial equations which are
satisfied between the joint parameters.  For example, $p_{\texttt{AAAA}} = p_{\texttt{CCCC}}$
holds for both JC69 and K81, but since this equation is true for all three trees,
will ignore it and similar equations.

\begin{example}\rm
	\label{ex:4pt}
Consider the four-point condition
on a tree metric \cite{Buneman1974}.  It says that if $d$ is a tree metric on $(ij : kl)$, then
\begin{equation}\label{eq:4pt}
d_{ij} + d_{kl} <  d_{ik} + d_{jl} = d_{il} + d_{jk}.
\end{equation}
Given a probability distribution $p$, the maximum likelihood Jukes-Cantor distance is
\begin{equation} \label{eq:JCdist}
d_{ij} = - \frac 3 4 \log \left( 1 - \frac {4 m_{ij}} {3}\right)
\end{equation}
where $m_{ij}$ is the fraction of mismatches between the two sequences, e.g.,
\[
m_{12} = \sum_{w,x,y,z \in \{A,C,G,T\}, w \neq x} p_{wxyz}.
\]
After substituting in (\ref{eq:4pt}) and exponentiating, the equality becomes
\begin{equation}\label{eq:4ptp}
\left(1 - \frac {4} {3} m_{ik}\right) \left(1 - \frac {4} {3} m_{jl}\right) =
\left(1 - \frac {4} {3} m_{il}\right) \left(1 - \frac {4} {3} m_{jk}\right). 
\end{equation}
This observation is originally due to Cavender and Felsenstein
\cite{Cavender1987}.  The difference of the two sides of (\ref{eq:4ptp}) is a
quadratic polynomial in the joint probabilities which we will call the
four-point polynomial. 
\end{example}

\section{Methods}
\label{sec:methods}
Since both models we consider are \emph{group-based}, it is easiest to work in Fourier
coordinates which can roughly be thought of as the $m_{ij}$ coordinates in
Example~\ref{ex:4pt} (cf.\ \cite{Hendy1989,Evans1993,Sturmfels2005}).  The
website \url{http://www.math.tamu.edu/~lgp/small-trees/} contains lists of
invariants for different models on trees with a small number of taxa.

Our first task is building a set of invariants for the two models.  The above
website shows 33 polynomials (plus two implied linear relations) for 
JC69 and 8002 polynomials for K81.
However, 
these sets of invariants 
are not closed under the symmetries of $T_1$.
That is, each tree can be
written in the plane in eight different ways (for example, the tree $\Ta$ can
be written as (01 : 23),  (10 : 23), \dots, (32 : 10)), and each of these induces a
different order on the probability coordinates $p_{ijkl}$.  We need a set of
invariants which does not change under this reordering if we don't want the
resulting algorithm  to depend on the order of the input sequences.

\begin{figure}
	\centering
	\includegraphics[width=.35\textwidth]{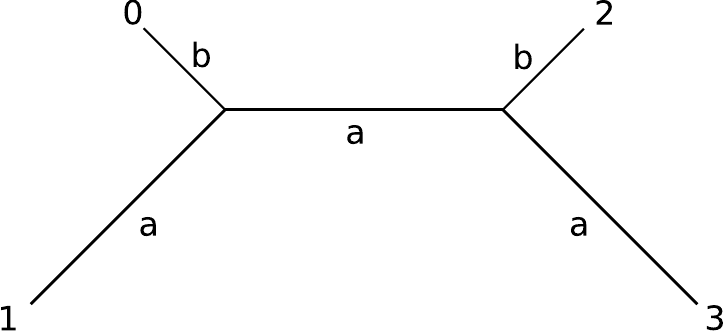} 
	\caption{The tree used in the simulations. Branch lengths $a$ and $b$
	ranged from $0.01$ to $0.75$ in intervals of $0.02$.}
	\label{fig:param}
\end{figure}

After performing this calculation, we are left with 49 polynomials for JC69 and
11612 for K81. 
However, our metric learning algorithms
run slowly as the number of invariants grows, so we had to find a subset of
a more manageable size.  
We cut down the K81 invariants by testing each of the
11612 invariants individually on the entire parameter space and only keeping
those which had good individual reconstruction rates.  
Specifically,  we picked several different values for
$\gamma, \alpha, \beta$ and kept only those invariants which gave over a 62\%
reconstruction rate individually for sequences of length 100.  The result of
this calculation is sets of invariants $\bof^{JC69}_i$ and $\bof^{K81}_i$ of
cardinality 49 and 52.

\begin{definition}
Given a probability distribution $\hat{p}$, and invariants $\bof_i = (f_{i,1},
\dots, f_{i,n})$, for tree $T_i$ (for $i = 1,2,3$), let 
\[
\bof_i(\hat{p}) = (f_{i,1}(\hat{p}), \dots, f_{i,n}(\hat{p}))
\]
be the point in $\rr^{n}$ obtained by evaluating the invariants for $T_i$ at
$\hat{p}$.  
\end{definition}

If the probability distribution $\hat{p}$ actually comes from the model $T_1$,
then we will have $\bof_1(\hat{p}) = 0$, $\bof_2(\hat{p}) \neq 0$, and $\bof_3(\hat{p})
\neq 0$ generically (that is, except for points $\hat{p}$ which lie on the
intersection of two or more models).  This fact suggests that we can just pick
the tree $T_i$ such that $\bof_i$ is closest to zero.  However, the next example
shows that it is quite important to pick good polynomials and weigh them properly.

\begin{example}\rm
Figure~\ref{fig:hist} shows the distribution of four of the invariants from $\bof^{JC}_i$
on data from simulations
of 1000 i.i.d.\
draws from the Jukes-Cantor model on $T_1$ over a varying set of parameters.
The histograms show the distributions for the simulated tree ($T_1$) in yellow
and the distributions for the other trees in gray and black.
\begin{figure}
	\centering
	\includegraphics[width=.5\textwidth]{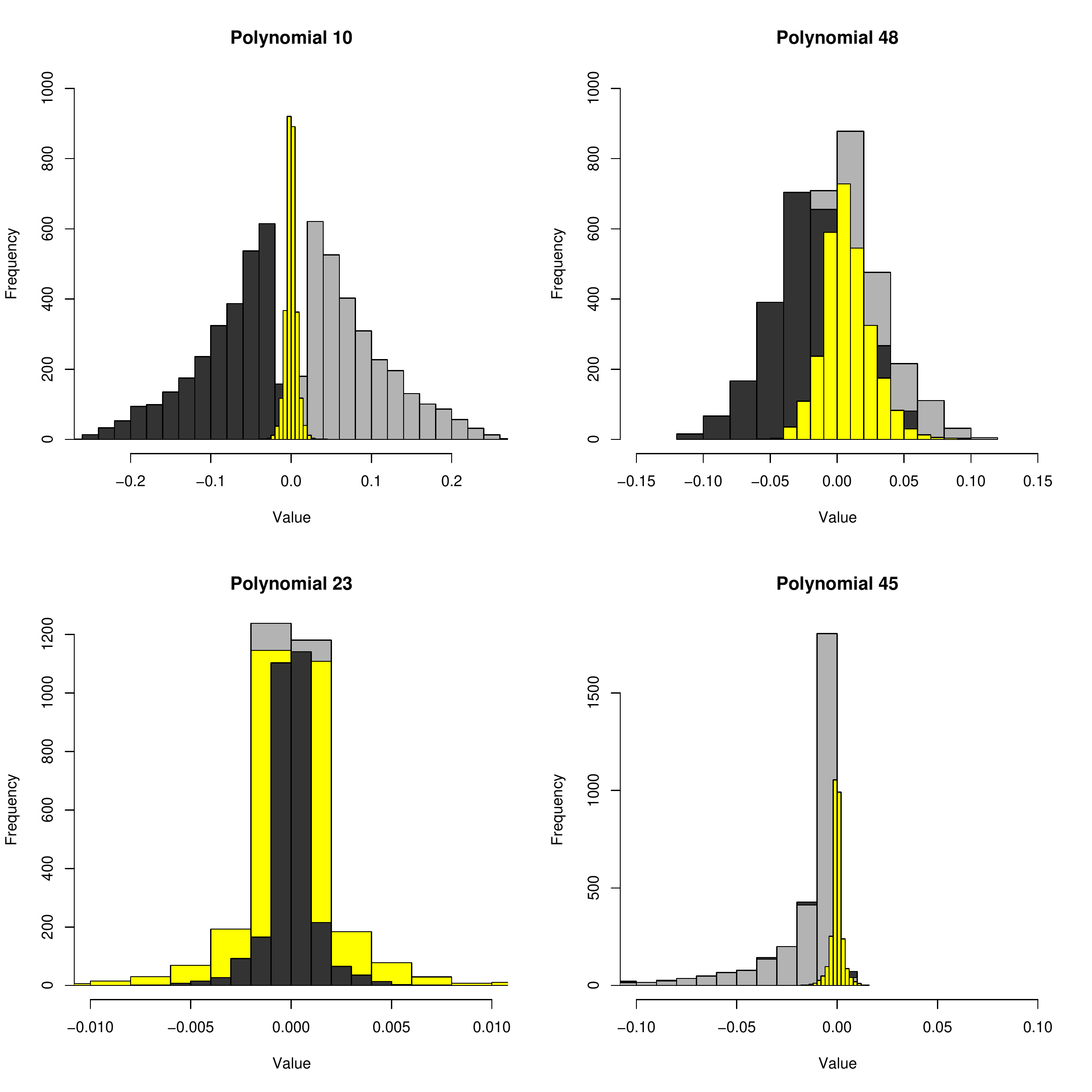}
	\caption{Distributions of four polynomials $f_{i,10}, f_{i,48}, f_{i,23}$ 
	and $f_{i,45}$ on simulated data.
	The yellow histogram corresponds to the correct tree, the black and gray are
	the other two trees.}
	\label{fig:hist}
\end{figure}

Polynomial 10 (upper left) distinguishes nicely between the three trees
with the correct tree tightly distributed around zero. It is correct 97\% of
the time on our space of trees (Figure~\ref{fig:param}). Polynomial 48 (upper right) also shows power to
distinguish between all three trees, but
the distributions are much more overlapping --- it is only correct 50.8\% of the time.
Polynomial 10 is the four-point invariant from Example~\ref{ex:4pt}, polynomial
48 is one of Lake's linear invariants.
The two other examples show a polynomial (23) which is biased towards selecting
the wrong tree (only 16\% correct), and a polynomial (45) for which the correct tree is tightly
clustered around zero, but the incorrect trees are indistinguishable and have wide
variance (88.9\% correct).


The parameters used for the simulations are described in
Figure~\ref{fig:param}.  Since 1000 samples should be quite enough to determine
the structure of a tree on four taxa, it is revealing that many of the
individual polynomials are quite poor 
(the mean prediction rate for all 49 polynomials is only 42\%).  
The invariants have quite different variances and means and it is not optimal
to take each one with equal weight.  
\end{example}


This example shows that we need to scale and weigh the individual invariants.
Recall that for a positive (semi)definite matrix $A$ 
the Mahalanobis (semi)norm $\|\cdot \|_A$ is defined by 
\[
\| x \|_A = \sqrt{x^t A x}.
\]
Notice that since $A$ is positive semidefinite, it can be written as $A = U D
U^t$ where $U$ is  orthogonal and $D$ is diagonal with non-negative entries.
Thus the square root $B = U \sqrt{D} U^t$ is unique.  Now since $\|x\|_A^2
= x^tAx = (Bx)^t(Bx)=\|Bx\|^2$, we can view learning such a metric as finding a
transformation of the space of invariants that replaces each point $x$ with $Bx$ under the 
Euclidean norm.
Accordingly, we will be searching for a positive semidefinite
matrix $A$ on the space of invariants which is ``optimal''.

Let
$\hpt$ be an empirical probability distribution generated from a phylogenetic
model on tree $T_1$ with parameters $\theta$.
We wish to find $A$ such that the condition
\[
\|\bof_1(\hpt))\|_A < \min\left( \|\bof_2(\hpt)\|_A, \|\bof_3(\hpt)\|_A \right)
\]
is typically true for most $\hpt$ chosen from a suitable parameter space $\Theta$.  

Now suppose that $\Theta$ is a finite set of parameters from which we generate
training data $\bof_1(\hpt), \bof_2(\hpt), \bof_3(\hpt)$ for $\theta \in \Theta$.
As we saw above, each of the eight possible ways of writing
each tree induces a signed permutation of the coordinates of each
$\bof_i(\hpt)$.  We write these permutations in matrix form as $\pi_1, \dots,
\pi_8$.
Given this training data, we wish to solve the following optimization problem.

\begin{alg}\rm
	\label{alg:meta}
	{\bf (Metric learning for invariants)}

	\noindent {\em Input:}  model invariants $\bof_i$ for $T_i$
	and a finite set $\Theta$ of model parameters.

	\noindent {\em Output:} a semidefinite matrix $A$ 

	\noindent {\em Procedure:}
	\begin{enumerate}
		\item Reduce the sets $\bof_i$ of invariants to a manageable size by testing individual invariants
			on data simulated from $\theta \in \Theta$.
		\item Augment the resulting sets so that they are closed under the eight
			permutations of the input which fix tree $T_1$.
		\item Compute the signed permutations $\pi_1, \dots, \pi_8$ which are induced on the
			invariants $\bof_1$ by the above permutations.
		\item Solve the following semidefinite programming problem:
			\begin{equation*}
				\begin{array}{ll}
					\text{Minimize: } &\sum_{\theta \in \Theta} \xi(\theta) + \lambda \tr A\\
					\text{Subject to: } \quad
					&\| X_1 (\hpt) \|_A^2   + \gamma \leq \min\left( \| X_2 (\hpt)\|_A^2 ,  \|X_3(\hpt) \|_A^2 \right)  + \xi(\theta),\\
					& \pi_i A = A \pi_i \quad\text{for } 1 \leq i \leq 8,\\
					&A \succeq 0, \quad\text{and}\\
					&\xi(\theta)  \geq 0,
				\end{array}
			\end{equation*} 
			where $A\succeq 0$ denotes that $A$ is a positive semidefinite
			matrix. 
		\item Alternatively, if we restrict $A$ to be
			diagonal, this becomes a linear program and can be solved for much
			larger sets of invariants and parameters.
	\end{enumerate}
\end{alg}

In the optimization step, we use a regularization parameter $\lambda$ to keep $A$
small and a margin parameter $\gamma$ to increase the margin between the
distributions.  This is a convex optimization problem with a linear objective
function and linear matrix equality and inequality constraints. Hence it is a
semidefinite programming (SDP) problem. 
The SDP problem above has a unique optimizer and can be solved in polynomial
time. Its complexity depends on the capacity of the set $\Theta$ since each
point in $\Theta$ contributes a linear constraint.

For the range of parameters we consider in Section~\ref{sec:results}, we use
$\#(\Theta)=1444$.  In our experiments to solve the SDP we use SeDuMi 1.1
\cite{Sturm1999} or DSDP5
\cite{BenYe05} with YALMIP \cite{Lofberg2004} as the parser.
Matlab code to implement the above algorithm can be found at
\url{http://math.stanford.edu/~yuany/metricPhylo/matlab/}. 

We found that although SeDuMi often runs into numerical issues, it generally
finds a good matrix $A$ with competitive performance to the neighbor-joining
algorithm. DSDP is better in dealing with numerical stability at the cost of
more computational time. We have found that setting $\lambda = 0.0001$ and
$\gamma = 0.005$ gives good results in our situation. For example, in the case
of the JC69 model with a $49\times49$ semidefinite matrix $A$, YALMIP-SeDuMi takes 55.7
minutes to parse the constraints and solve the SDP, while YALMIP-DSDP takes
167.1 minutes to finish the same job.  For details on experiments, see the next
section. 

Our algorithm was inspired by some early results on metric learning algorithms
such as \cite{XingNg03} and \cite{ShaSinNg04}, which aim to find a
(pseudo)-metric such that the mutual distances between similar examples are
minimized while the distances across dissimilar examples or classes are kept
large. Direct application of such an algorithm is not quite suitable in our setting. 
As shown in Figure~\ref{fig:hist},
the correct tree points are overlapped by the two incorrect trees. For
points in the overlapping region, it is hard to tell whether to shrink or
stretch their mutual distance. However, when the points appear in triples, it
is possible that for each triple the one closest to zero is generated from the
correct tree. Our algorithm is based on such an intuition and proved successful
in experiments. 

After using Algorithm~\ref{alg:meta} to find a good metric $A$, the following
simple algorithm allows us to construct trees on four taxa.

\begin{alg}\rm
\label{alg:1}
{\bf (Tree construction with invariants)}  

\noindent {\em Input:} 
A multiple alignment of 4 species and a semidefinite matrix $A$ from Algorithm~\ref{alg:meta}.

\noindent {\em Output:}
A phylogenetic tree on the 4 species (without branch lengths).
 
\noindent {\em Procedure:}
\begin{enumerate}
	\item Form empirical distributions $\hat{p}$ by counting columns of the alignment.
	\item Form the vectors $\bof_i = (f_{i,1}(\hat{p}), \dots, f_{i,n}(\hat{p}))$ for $1 \leq i \leq 3$.
	\item Return $T_i$ where 
		the vector $\bof_i$ has smallest $A$-norm $\|\bof_i\|_A = \sqrt{\bof_i^t A \bof_i}$.
\end{enumerate}
\end{alg}


\section{Results}
\label{sec:results}

We tested our metric learning algorithms for the invariants $\bof^{JC69}$ and
$\bof^{K81}$ as described above.  
We trained two metrics for JC69 on 
the tree in Figure~\ref{fig:param}.
The first used simulations from branch lengths between $0.01$ and $0.75$ on a
grid with increments of magnitude $0.02$, for a total of $1444$ different
parameters.  
This region of parameter space was chosen for direct comparison with
\cite{Huelsenbeck1995,Casanellas2006}.
The second metric used parameters $0.01 \leq a \leq 0.25$ and $0.51 \leq b \leq
0.75$ with increments of $0.01$, giving $625$ trees with very short interior edges.
Similarly for K81, we learned a metric using parameters $0.01 \leq a,b \leq 0.75$ and
several sets of $\gamma, \alpha, \beta$.
After learning metrics, we performed simulation tests on the same space of
parameters that was used to train.  We compared neighbor-joining
\cite{Saitou1987}, phylogenetic invariants with the $l_1$ or $l_2$ norm and
phylogenetic invariants with our learned norms. 

\begin{figure}
	\centering
	\begin{tabular}{cc}
	\includegraphics[height=.35\textheight]{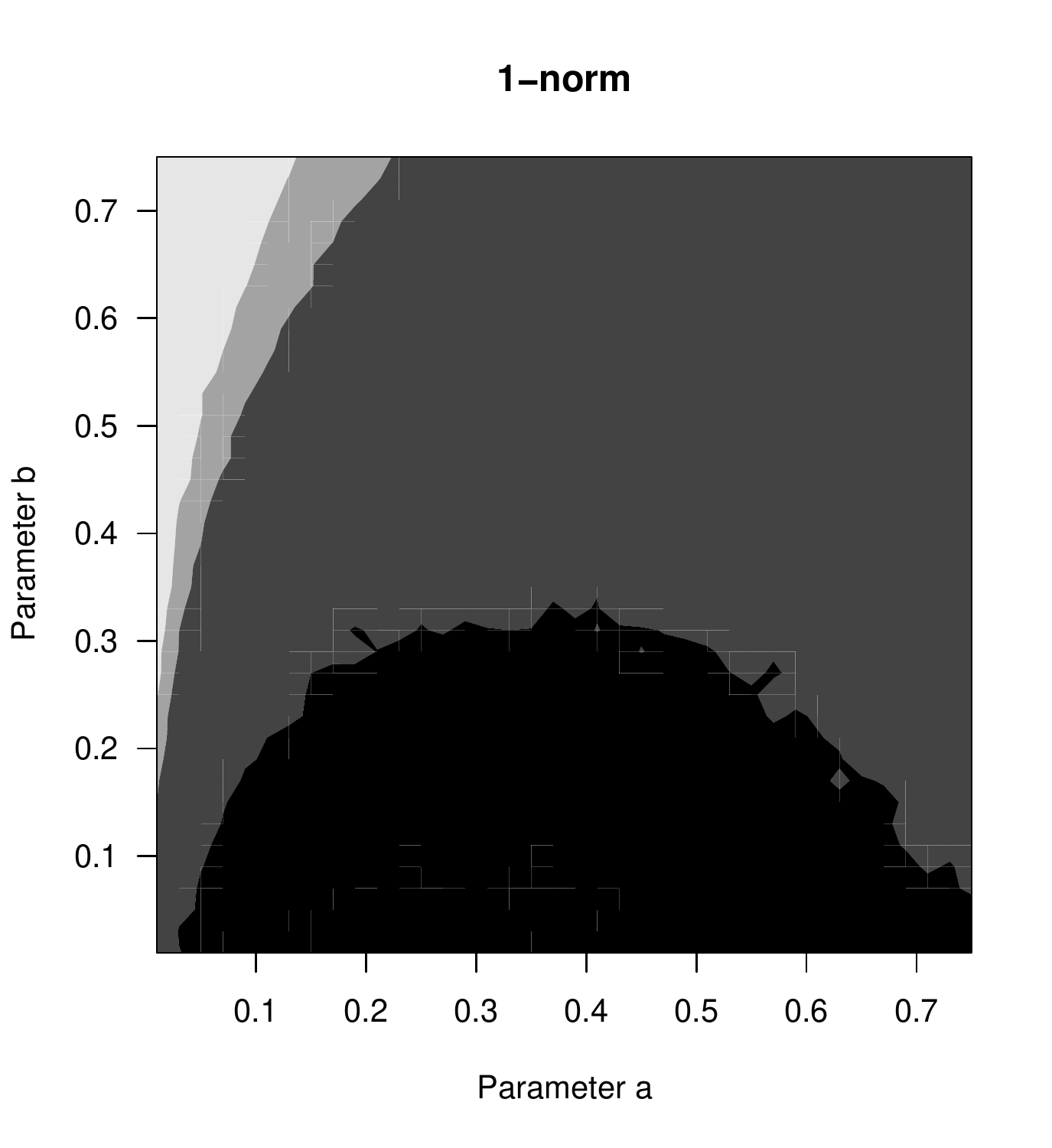} & 
	\includegraphics[height=.35\textheight]{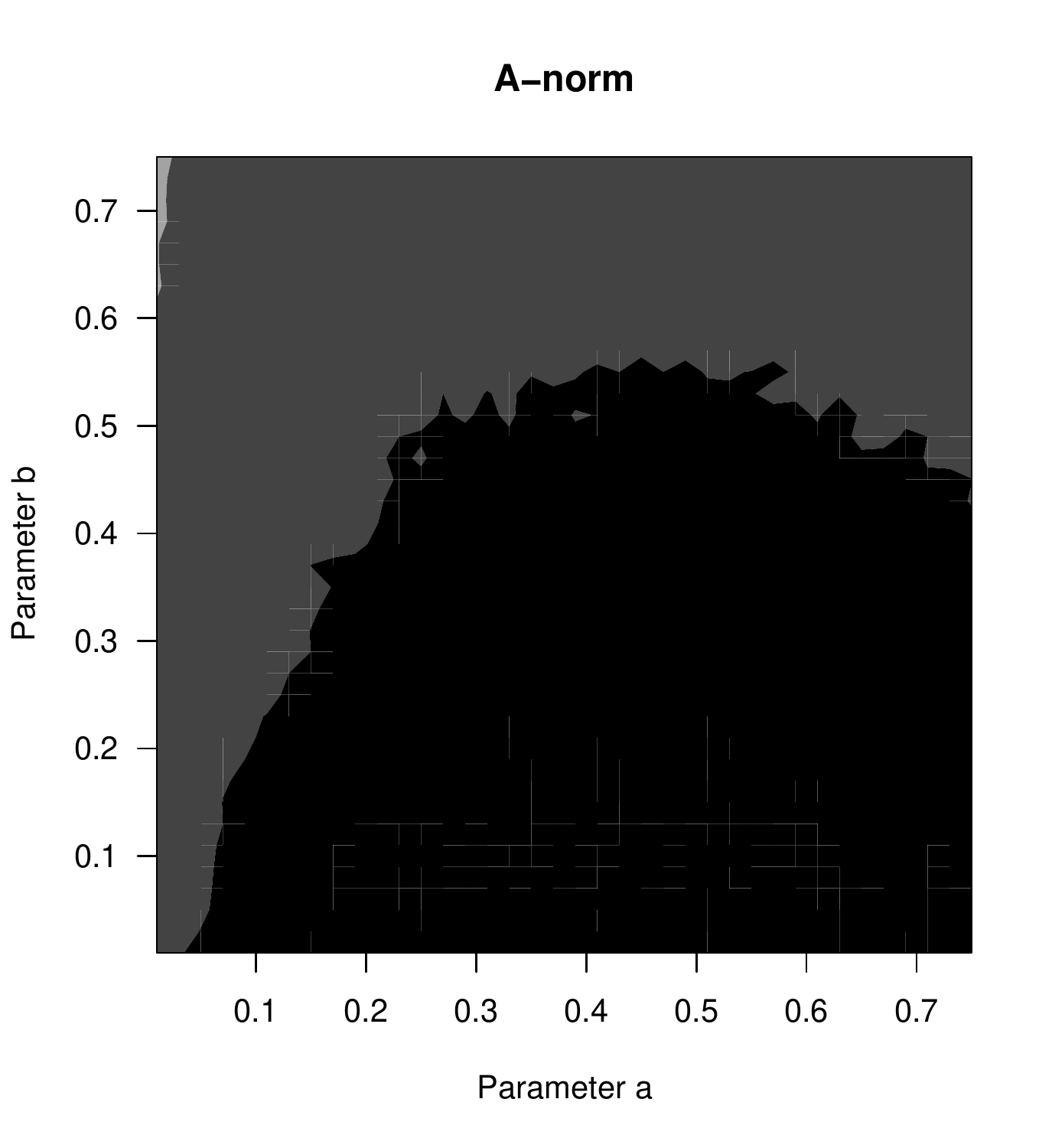} \\
	\includegraphics[height=.35\textheight]{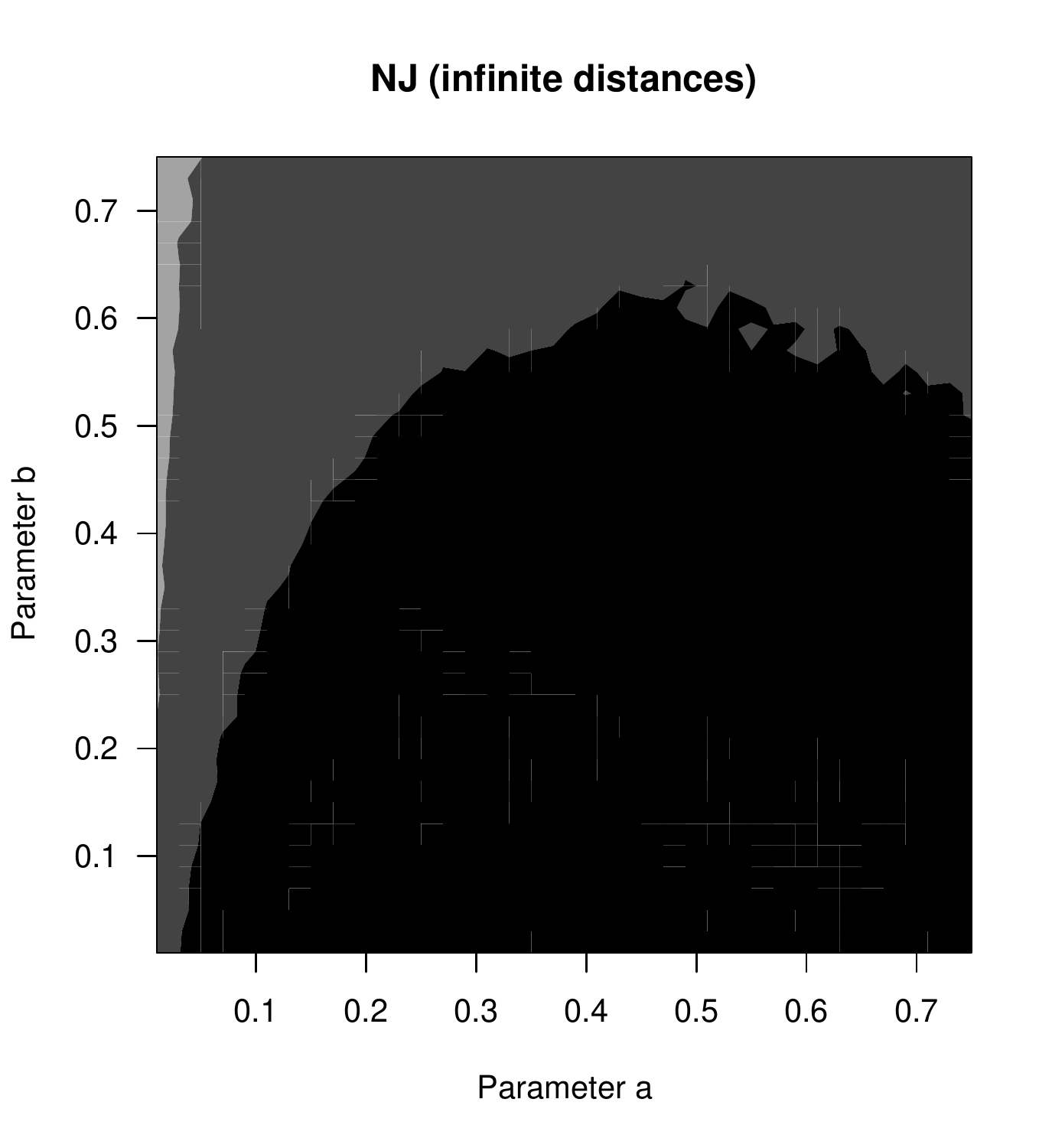} &
	\includegraphics[height=.35\textheight]{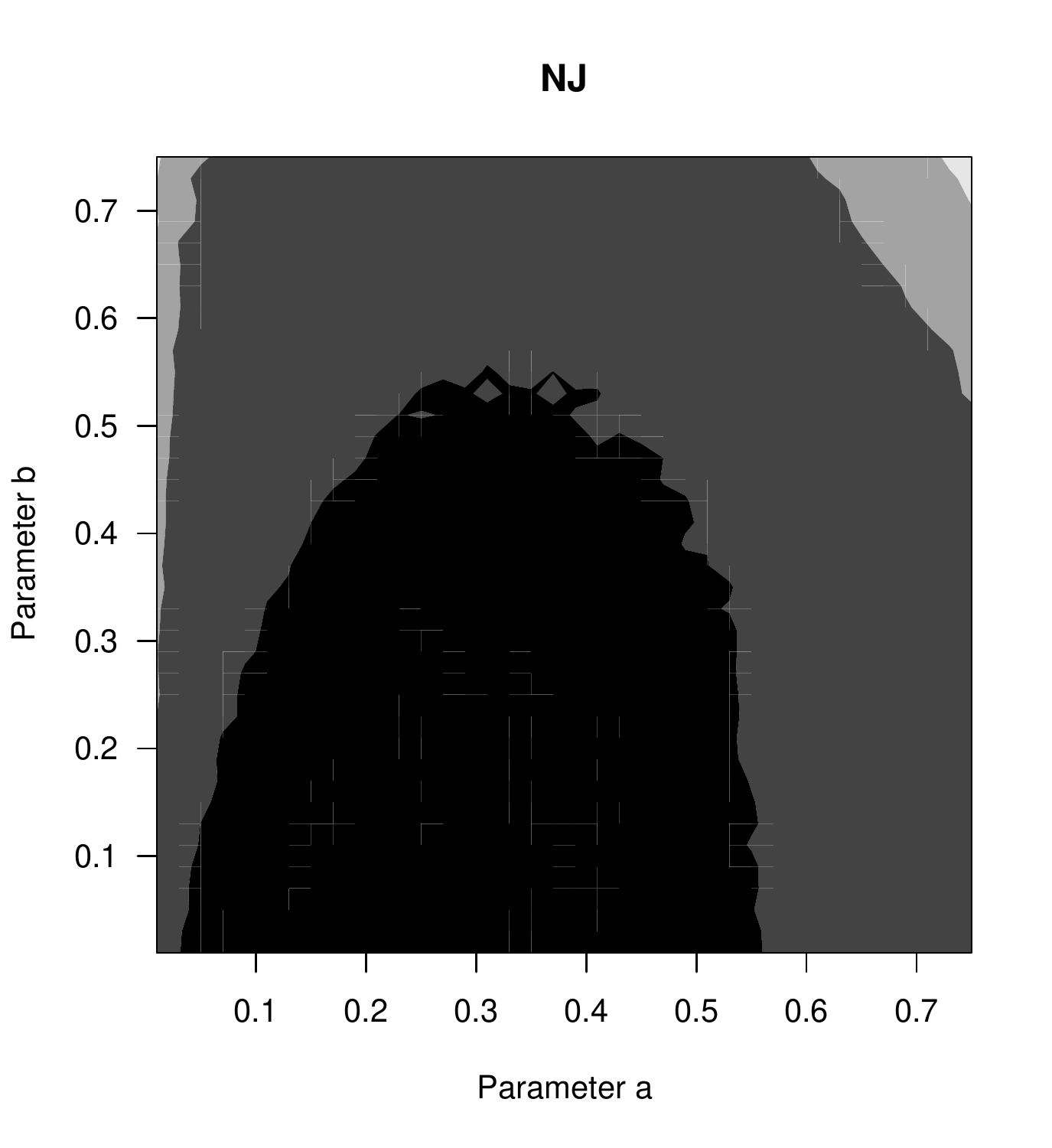} \\
\end{tabular}
	\caption{Contour plots for the three reconstruction methods for the
	Jukes-Cantor model over parameter space with alignments of length 100.
	Black areas correspond to parameters
	$(a,b)$ for which the tree was reconstructed correctly  over 95\% of the
	time, gray for over 50\%, light gray for over 33\%, and white for under 33\%.  } 
	\label{fig:contour}
\end{figure}

Since the edge lengths are large for part of the parameter space, we often see
simulated alignments with more than 75\% mismatches between pairs of taxa. In
such a case, (\ref{eq:JCdist}) returns infinite distance estimates under the
Jukes-Cantor
model.  So the results depend on how neighbor joining treats
infinite distances.  In PHYLIP
\cite{PHYLIP}, the program \texttt{dnadist} doesn't return a distance
matrix if some distances are infinite.  However, in PAUP*, infinite distances are set to a fixed large
number. 
Since we are only concerned with the tree topology, we believe that the most
fair comparison is between phylogenetic invariants and the method of PAUP*.
However, it should be noted that this can make a major difference in results using neighbor joining, since often the correct tree can be returned even if some distances are infinite. See Figure~\ref{fig:contour} for an example of the difference and be warned that comparison between simulations studies done in different ways is difficult.

Table~\ref{tab:JC} shows the results of 100 simulations at each of the 1444
parameter values for various sequence lengths using the JC69 model.  It gives
the percent correct over all 144,400 trials for
five different methods: invariants with $l_1$, $l_2$, and $A$-norms and
neighbor joining (using Jukes-Cantor distances and allowing infinite distances).  The contour
plots in Figure~\ref{fig:contour} show how the reconstruction rates vary across
parameter space for the five methods for a sequence length of 100.  
Notice that the $A$-norm shows
particularly good behavior over the entire range of parameters, even in the
``Felsenstein zone'' in the upper left corner.  
When trained on the Felsenstein zone, the learned metric can perform even
better.  Table~\ref{tab:JC} shows the result of training a metric on this zone.
Notice that the $A$-norm is now quite a bit better than neighbor joining, even
though the $l_1$ and $l_2$ norms are terrible.  However, this learned norm is slightly worse on the whole parameter space than the metric trained on the whole space.

\begin{table}
	\centering
	\begin{tabular}{cccccl@{\hspace{1cm}}ccccc}
		\multicolumn{5}{c}{Full parameter space} & &
		\multicolumn{5}{c}{Felsenstein zone}\\
		Length & $l_1$  & $l_2$ & $A$-norm  &  NJ & &
		Length & $l_1$  & $l_2$ & $A$-norm  &  NJ\\
		\cline{1-5} \cline{7-11} 
		 25 & 62.7  & 59.8  & 74.5  & 75.5 &&   25 & 31.8 & 31.3 & 58.9 & 52.5\\ 
		 50 & 71.9  & 66.3  & 85.0  & 85.9 &&   50 & 35.9 & 33.0 & 69.9 & 63.2\\
		 75 & 76.7  & 69.6  & 90.0  & 90.4 &&   75 & 39.2 & 34.4 & 76.5 & 69.5\\
		100 & 79.8  & 72.0  & 92.7  & 92.9 &&  100 & 42.2 & 35.8 & 81.2 & 73.9\\
		200 & 86.4  & 77.6  & 97.0  & 96.6 &&  200 & 50.9 & 39.1 & 90.1 & 83.5\\
		300 & 89.2  & 80.1  & 98.2  & 97.7 &&  300 & 55.4 & 40.4 & 93.6 & 87.8\\
		400 & 91.1  & 82.1  & 98.7  & 98.2 &&  400 & 59.5 & 41.4 & 95.1 & 90.2\\
		500 & 92.3  & 83.5  & 99.0  & 98.4 &&  500 & 62.4 & 42.5 & 95.6 & 91.4\\
		\cline{1-5} \cline{7-11} 
	\end{tabular}
	\caption{Percent of trials reconstructed correctly for the Jukes-Cantor
	model over the entire parameter space and the Felsenstein zone for the
	respective metrics.}
	\label{tab:JC}
\end{table}

Table~\ref{tab:K81} shows results for the K81 model under two choices of $(\gamma, \alpha, \beta)$.
We only report the $l_2$ scores, since the $l_1$ scores are similar.  Of note
is the column ``$l_2$ restrict'' which shows the $l_2$ norm on the top $52$
invariants as ranked by individual power on simulations as in the previous
section.  This column is better than the $l_2$ norm on all $11612$ invariants,
showing that many invariants are actually harmful.
The $A$-norm again improves on even the restricted $l_2$ and beats
neighbor-joining (run with K81 distances) on all examples.

\begin{table}
	\centering
	\begin{tabular}{cccccl@{\hspace{1cm}}ccccc}
		\multicolumn{5}{c}{$(\gamma, \alpha, \beta) = (0.1, 3.0, 0.5)$}&&
		\multicolumn{5}{c}{$(\gamma, \alpha, \beta) = (0.2, 0.5, 0.3)$}\\
length & $l_2$ & $l_2$ restrict & A & NJ&&
length & $l_2$ & $l_2$ restrict & A & NJ\\
\cline{1-5} \cline{7-11}
 25 & 59.2 & 66.9 & 71.5 & 62.7 && 25 & 63.7 & 67.4 & 70.9 & 65.0 \\ 
 50 & 68.0 & 77.5 & 82.1 & 72.7 && 50 & 71.6 & 77.5 & 81.1 & 74.2 \\
 75 & 73.4 & 82.7 & 86.9 & 79.3 && 75 & 75.4 & 82.7 & 86.4 & 80.5 \\
100 & 76.8 & 85.8 & 89.7 & 82.6 &&100 & 77.6 & 85.8 & 89.3 & 83.4 \\
200 & 84.6 & 90.8 & 94.3 & 90.1 &&200 & 83.3 & 91.3 & 94.1 & 89.7 \\
300 & 88.2 & 92.6 & 93.1 & 95.6 &&300 & 86.4 & 93.2 & 95.7 & 91.8 \\
400 & 90.1 & 93.6 & 96.4 & 94.8 &&400 & 88.5 & 94.3 & 96.5 & 93.0 \\
500 & 91.4 & 94.2 & 96.9 & 95.7 &&500 & 90.0 & 93.7 & 96.3 & 93.2 \\
\cline{1-5} \cline{7-11}
	\end{tabular}
	\caption{Percent of trials reconstructed correctly 
	for the Kimura 3-parameter model over the entire parameter space for two choices of $\gamma, \alpha, \beta$.} 
	\label{tab:K81}
\end{table}

\section{Discussion}
\label{sec:dis}

We have shown that machine learning algorithms can substantially improve the
tree construction performance of phylogenetic invariants.  As an example, for
sequences of length 100, the four-point invariant (Example~\ref{ex:4pt}) for
the K81 model is correct 82\%  of the time on data simulated from K81 with
parameters $(0.1, 3.0, 0.5)$.  This is quite a bit better than the 
$l_2$ norm on all 11612 invariants (76.8\%, Table~\ref{tab:K81}).  

The paper \cite{Casanellas2007} describes an algebraic method for picking a
subset of invariants for the K81 model.  They reduce to 48 invariants which give
an improvement over all 11612 invariants (up to 82.6\% on the above
example using the $l_2$ norm).  However, of these 48, only 4 of them are among
the top 52 we selected for $\bof^{K81}$, and the remaining 44 invariants are
mostly quite poor (42\% average accuracy). After taking the closure of these 48 invariants, there are 156
total and the performance actually drops to 78.3\%.  It seems that the
conditions for an invariant to be powerful are not particularly related to the
algebraic criterion used in \cite{Casanellas2007}.

All invariant based methods heavily depend on the set of invariants that we
begin with.  Learning diagonal matrices $A$ had mixed performance, which further
suggests that the generating set we are using for the invariants is
non-optimal.  We believe that it is an important mathematical problem to
understand what properties are shared by the good invariants.  We suggest that
symmetry might be an important criterion to construct other polynomials like
the four-point condition with good power.

The learned metrics in this paper are somewhat dependant on the
parameters chosen to train them.  This can be a benefit, as it allows us to
train tree construction algorithms for specific regions of parameter space
(e.g., the Felsenstein zone).  However, we hope that improvements to the metric
programming will allow us to train on larger parameter sets and thus obtain
uniformly better algorithms.

Notice that these methods only recover the tree topology, not the edge lengths.
We believe that if the edge lengths are needed, they should be estimated after
building the tree, in which case standard statistical methods such as maximum
likelihood can be used easily.
While the invariants discussed in this paper may not be practical for large
trees, we believe there is great use in understanding fully the problem of
building trees on four taxa.  For example, these methods can either be used
as an input to quartet-based tree construction algorithms or as a verification
step for larger phylogenetic trees.

For a method of building trees on more than four taxa using phylogenetic
invariants, see \cite{Eriksson2005b,Kim2006}, which use numerical linear
algebra to evaluate invariants given by rank conditions on certain matrices in
order to construct phylogenetic trees.  This amounts to evaluating many
polynomials at once, allowing it to run in polynomial time.

The matrices $A$ for JC69 and K81 used in the tests can be found at
\url{http://stanford.edu/~nke/data/metricPhylo}.  A software package that can
run these tests is available at the same website.  It includes a program for
simulating evolution using any Markov model on a tree and several programs
using phylogenetic invariants.

\section*{Acknowledgments}
N.~Eriksson was supported by NSF grant DMS-0603448 and wishes to thank MSRI and
the IMA for their hospitality. Y.~Yao was supported by DARPA grant 1092228.  We
wish to thank E.\ Allman, M.\ Drton, D.\ Ge, F.\ Memoli, L.\ Pachter, J.\
Rhodes, and Y.\ Ye for helpful comments and especially G.\ Carlsson for his
encouragement and computational resources. 

\bibliographystyle{alpha}

\end{document}